\documentstyle[multicol,aps]{revtex}
\begin{document}
\title{Role of break-up processes in fusion enhancement 
 of drip-line nuclei\\at energies below the Coulomb barrier} 

\author {K. Hagino,$^1$ A. Vitturi,$^2$ C.H. Dasso,$^3$
and  S.M. Lenzi$^2$}

\address{$^1$Institute for Nuclear Theory, 
Department of Physics, University of Washington, 
Seattle, WA98195}

\address {$^2$Dipartimento di Fisica, Universit\`a di Padova and INFN, 
   Padova, Italy}

\address{$^3$Departamento de F\'{\i}sica At\'omica, Molecular y
Nuclear, Universidad de Sevilla, Spain}

\date{\today}

\maketitle

\begin{abstract}
 We carry out realistic coupled-channels calculations for 
 $^{11}$Be + $^{208}$Pb reaction in order to discuss 
 the effects of break-up of the projectile nucleus 
 on sub-barrier fusion. 
 We discretize in energy the particle continuum states, 
 which are associated with the break-up process, 
 and construct the coupling form factors to these states 
 on a microscopic basis. 
 The incoming boundary condition is 
 employed in solving coupled-channels equations, which 
 enables us to define the flux for complete fusion inside 
 the Coulomb barrier. It is shown that complete fusion 
 cross sections are significantly enhanced 
 due to the couplings to the continuum states compared with 
 the no coupling case at energies below the Coulomb barrier,  
 while they are hindered at above barrier energies. 

\end{abstract}


\begin{multicols}{2}

 Quantum tunneling in systems with many degrees of freedom \cite{CL81} 
 has attracted much interest in recent years in many fields of 
 physics and chemistry \cite{JJAP93}. 
 In nuclear physics, heavy-ion fusion reactions at energies near 
 and below the Coulomb barrier are typical examples for this 
 phenomenon. 
 In order for fusion processes to take place, the Coulomb barrier 
 created by the cancellation between the repulsive Coulomb force 
 and the attractive nuclear interaction has to be overcome. 
 It has by now been well established that the coupling of the relative 
 motion of the colliding nuclei to nuclear intrinsic excitations as well 
 as to transfer reaction channels cause large
 enhancements of the fusion cross section at subbarrier 
 energies over the predictions of a simple barrier penetration 
 model\cite{review}. 

 The effect of break-up processes on fusion, on the other hand, has 
 not yet been understood very well, and many questions have been raised 
 during the last few years both from the 
experimental~\cite{R98,T97,D99,S98,K98} 
 and theoretical~\cite{2,3,4,Y97} points of view.  
 The issue has become especially relevant in recent years due to the 
 increasing availability of radioactive beams.  
 These often involve weakly-bound systems close to the drip lines for
 which the possibility of projectile dissociation prior to or at the 
 point of contact cannot be ignored.
 Different theoretical approaches to the problem have led to 
 controversial results, not only quantitatively but also qualitatively. 
 The probability for fusion at energies below the barrier has been in 
 fact predicted to be either reduced~\cite{2,3} or enhanced~\cite{4,Y97} 
 by the coupling to the continuum states. 

 These investigations, however, were not satisfactory in view 
 of the rather simplified assumptions used in the treatment of both 
 the structure and reaction aspects of the problem.
 In refs.~\cite{2,3} the coupling to the break-up channels 
 was incorporated in terms of a ``survival factor'', a procedure that 
 underestimates the effects of the coupling in the classically forbidden 
 region, i.e. the dynamical modulation of the effective potential which 
 is most relevant at energies below the barrier. 
 Ref. \cite{4} took this effect into account but the entire continuum 
 space was mocked up by a single discrete configuration and an arbitrary 
 function was introduced to parametrize the radial dependence of the 
 couplings to such state.  

 In this letter we address the problem without resorting to 
 these approximations. 
 Realistic form factors to the continuum states are constructed by folding 
 the external nuclear and Coulomb fields with the
 proper single-particle transition densities, 
 obtained by promoting the
 last weakly-bound nucleon to the continuum states\cite{5}. 
 The reaction mechanism
 is described within a full quantal coupled-channel description\cite{6}. 
 The flux for complete fusion is separated inside the Coulomb 
 barrier from that for incomplete fusion 
 using the incoming boundary conditions \cite{HT98}.
 In order to isolate the genuine effect of the break-up process, 
 we include only the continuum states in the coupling scheme, 
 neglecting continuum-continuum coupling as well as other inelastic 
 channels such as bound excited states 
 in either reaction partner.
 For the same reason, we do not take into account static 
 modifications on the ion-ion potential which may arise from the halo 
 properties of the projectile\cite{TS91}.  

 As an example for our study we choose the fusion reaction 
 $^{11}$Be + $^{208}$Pb, where the projectile is generally regarded 
 as a good example of a system with a single neutron ``halo''.  
 In a pure single-particle picture, the last neutron in $^{11}$Be
 occupies the $2s_{1/2}$ state, bound by about 500 keV.
 The strong concentration of strength at the continuum threshold evidenced in
 break-up reactions\cite{7} has been mainly ascribed to the promotion of this
 last bound neutron to the continuum of $p_{3/2}$ and $p_{1/2}$ states
 at energy $E_c$ via the dipole field\cite{7,8}. 
 Since the presence of the first excited $1p_{1/2}$
 state (still bound by about 180 keV) may perturb the transition to the 
 corresponding $p_{1/2}$ states in the continuum\cite{9}, we prefer 
 here to consider only the contribution to the $p_{3/2}$ states.
 The initial $2s_{1/2}$ bound state and the continuum $p_{3/2}$ states 
 are generated by Woods-Saxon single-particle potentials whose depths
 have been adjusted to reproduce the correct binding energies for the
 $1p_{3/2}$ and $2s_{1/2}$ bound states.  In particular, one needs for the
 latter case a potential which is much deeper than the ``standard'' one.
 The form factor $F(r;E_c)$ as a function of the internuclear separation $r$
 and of the energy $E_c$ in the continuum is then obtained by folding the
 corresponding transition density with the external field generated by the
 target.  
 In addition to the Coulomb field, a Woods-Saxon nucleon-nucleus potential  
 is used, with parameters of $R=r_\circ A_{_T}^{1/3}$, $r_\circ=1.27$ fm, 
 $a=0.67$ fm, $V=(-51+33~(N-Z)/A)$ MeV, and $V_{ls}=-0.44~V$.  

 The dipole form factor $F(r;E_c)$ thus constructed 
 at several values of $r$ and $E_c$ 
 are shown in Fig. 1.  In Fig. 1(a), 
 we display the form factor as a function of $r$ for a fixed value of the 
 energy in the continuum ($E_c$ being 0.9 MeV).  
 Note the long tail of the nuclear contribution
 as a consequence of the large radial extension of the weakly-bound wave 
 function, resulting in the predominance of the nuclear form factor up
 to the unusual distance of about 25 fm.  The same reason gives rise to a 
 deviation of the Coulomb part from the asymptotic behavior 
 proportional to $r^{-2}$. 
 Note also the constructive interference of the nuclear and Coulomb parts,
 due to the negative $E1$ effective charge of the neutron excitation.
 In figs. 1(b) and 1(c), we 
 show, instead, the energy dependence of the form factors for a 
 fixed value of $r$.  
 At large values of $r$ the curves are peaked at very low energies,
 reflecting the corresponding behavior of the $B(E1)$.
 At distances around the barrier (which are most relevant to the fusion 
 process) the peaks of the distributions move to higher energies,  
 especially for the nuclear part. 
  
 In order to perform the coupled-channel calculation the distribution of 
 continuum states is discretized in bins of energy, associating
 to each bin the form factor corresponding to its central energy. We have
 considered the continuum distribution up to 2 MeV, with a step of 200 keV.
 In this way, the calculations are performed with 10 effective
 excited channels.  The ion-ion potential is assumed to have a Woods-Saxon 
 form with parameters $V_\circ= -$152.5 MeV, $r_\circ$=1.1 fm and 
 $a=$0.63 fm, a set that leads to the same barrier height as the 
 Aky\"uz-Winther potential. 
 At the distance inside the Coulomb barrier where the incoming boundary
 conditions are imposed, the flux for the entrance channel and 
 the excited break-up channels are evaluated separately. 
 Cross sections for complete fusion, leading to $^{219}$Rn, are then 
 defined using only the flux for the entrance channel, while those for 
 incomplete fusion which leads to $^{218}$Rn are defined 
 in terms of the flux for the break-up channels \cite{HT98}. 

 Figure 2(a) shows the results of our calculations. 
 The solid line represents the complete fusion cross section, while the 
 dashed line denotes the sum of the complete and incomplete fusion 
 cross sections. Also shown for comparison, by the thin solid line, is 
 the cross section in the absence of the couplings to the continuum
 states. 
 One can see that these enhance the fusion cross sections 
 at energies below the barrier over the predictions of 
 a one-dimensional barrier penetration model.
 Note that this is the case not only for the 
 total (complete plus incomplete) fusion probability, but also for the 
 complete fusion in the entrance channel.  
 This finding is qualitatively the same as that of Ref. \cite{Y97} 
 which used a three-body model to reach to this conclusion, 
 and supports the results of the original calculation performed in 
 Ref. \cite{4}. 
 As it has been emphasized there, accounting properly for the dynamical 
 effects of the coupling in the classically forbidden region is essential
 to arrive to this conclusion.

 The situation is completely reversed at energies above the barrier, 
 where fusion in the break-up channel becomes more
 important and dominates at the expense of the complete fusion. 
 As a consequence, the cross sections for complete fusion are 
 hindered when compared with the no-coupling case. 
 It is interesting to check if this hindrance is caused mainly 
 by the Coulomb interaction. 
 Figure 2(b) shows the effect of the individual 
 nuclear and Coulomb excitations separately. 
 It is apparent from this figure that 
 the nuclear coupling plays a more important role than the Coulomb one 
 in fusion reactions at energies below the Coulomb barrier. 
 This can be understood by considering that the nuclear process is 
 essentially dominated by the values of the potentials and couplings 
 around the Coulomb barrier.
 At energies above the barrier, on the other hand, both the Coulomb 
 and the nuclear break-up processes 
 play an important role in suppressing the fusion 
 cross sections. This is a characteristic feature of loosely bound 
 systems, where the nuclear form factor extends outside the Coulomb barrier. 
 For fusion of stable nuclei, the Coulomb break-up would become more important 
 in complete fusion at energies above the barrier.

 In summary, we have performed exact coupled-channels calculations for 
 weakly-bound systems using 
 realistic coupling form factors to discuss effects of break-up on 
 subbarrier fusion reactions. 
 As an example, we have considered the fusion of
 $^{11}$Be with a $^{208}$Pb target, taking into account the dipole 
 transition of the weakly-bound $2s_{1/2}$ neutron to the 
 $p_{3/2}$ continuum, which gives the dominant
 contribution to the low-energy $B(E1)$ response. 
 Couplings to bound excited states both in the projectile and 
 in the target nuclei as well as the static change of the ion-ion potential 
 were left aside in order to investigate genuine effects of the break-up 
 processes on fusion reactions. 
 We find that the coupling to break-up channels enhances cross sections 
 for the complete fusion at energies below the Coulomb barrier, while it 
 reduces them at energies above.

 Very recently, a complete fusion excitation function was measured for 
 the $^9$Be + $^{208}$Pb reaction at near-barrier energies by 
 Dasgupta {\it et al}. \cite{D99}. They showed that cross sections for 
 complete fusion are considerably smaller at above-barrier energies 
 compared with a theoretical calculation that reproduces the total 
 fusion cross section, in general agreement with our results. 
 We note, however, that it is not at all easy to identify
 reliable reference measurements to compare with at energies 
below the barrier. This feature may make it quite difficult to settle 
the issue on a purely experimental basis.

 The authors are grateful to the ECT* in Trento,
 the INT at the University of Washington (A.V.), 
 and the INFN Padova (K.H.) for their hospitality 
 and for partial support for this project.

\narrowtext
      
\begin{figure}
\caption{Coupling form factor $F(r;E_c)$ associated with the 
 dipole transition in
 $^{11}$Be from the neutron bound state $2s_{1/2}$ ($E_b$ = $-$500 keV)
 to the continuum state $p_{3/2}$ with energy $E_c$.  
 In (a) the Coulomb (dot-dashed line), nuclear (dashed) and 
 total (solid) 
 form factors are shown as a function of $r$ at the continuum energy 
 $E_c=0.9$ MeV.  In (b) and (c) the form factors are shown as a function
 of the energy $E_c$ in the continuum for $r=r_{barrier}$=11.6 fm and $r=30$ 
 fm, respectively.}  
\end{figure}
              
\begin{figure}
\caption{(a) Fusion cross section for the reaction $^{11}$Be + $^{208}$Pb 
 as a function of the bombarding energy in the center of mass frame. 
 The thin solid curve shows the results of the one-dimensional barrier 
 penetration as a reference.  
 The solid and the dashed lines are solutions of the coupled-channels 
 equations for the complete fusion and the complete plus incomplete fusion, 
 respectively. 
 (b) Complete fusion cross sections obtained by including 
 only the nuclear (dashed) and the Coulomb (dot-dashed) couplings, 
 in comparison with the case where both the couplings are considered (solid).}
\end{figure}
\end{multicols}

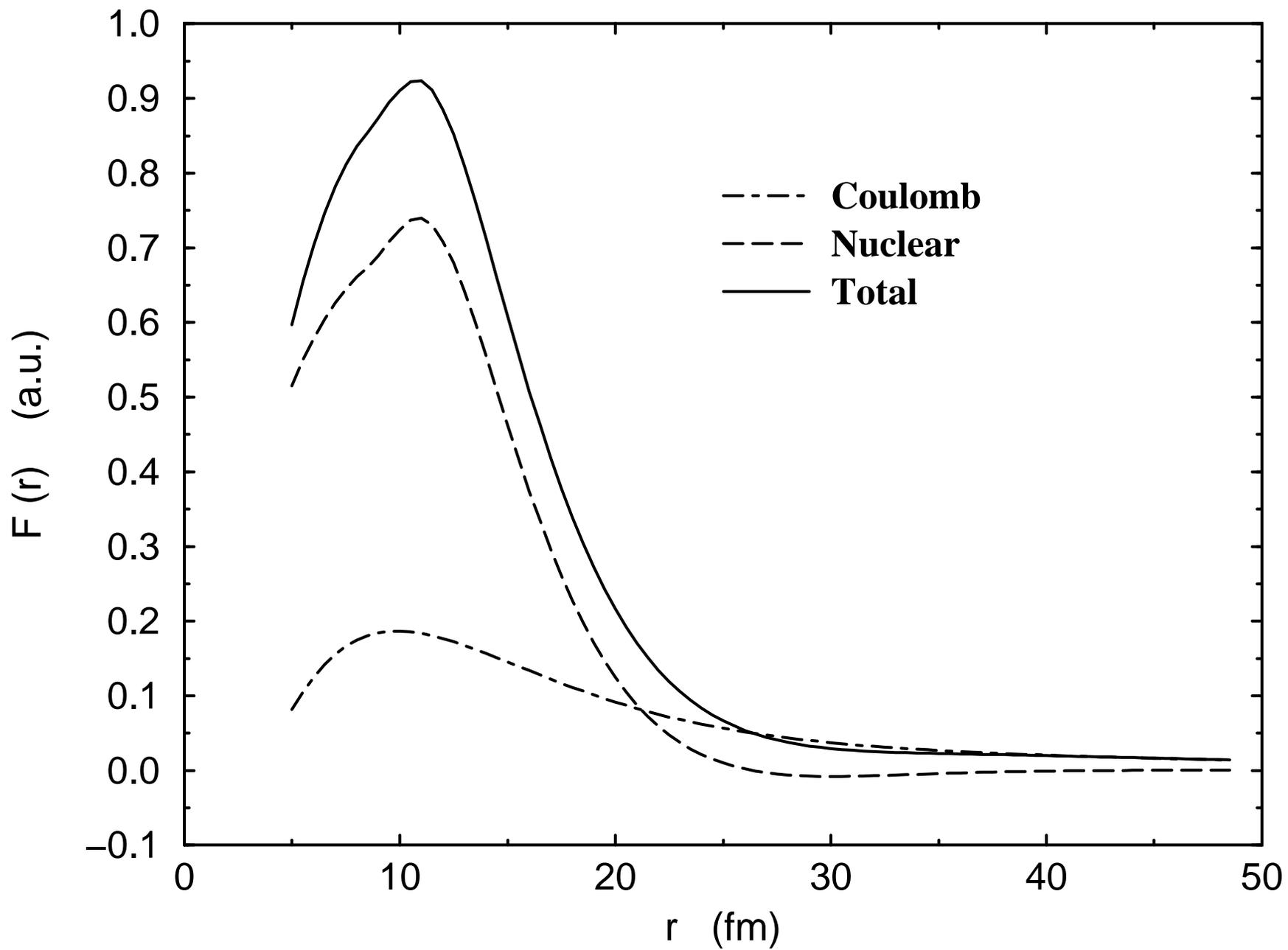
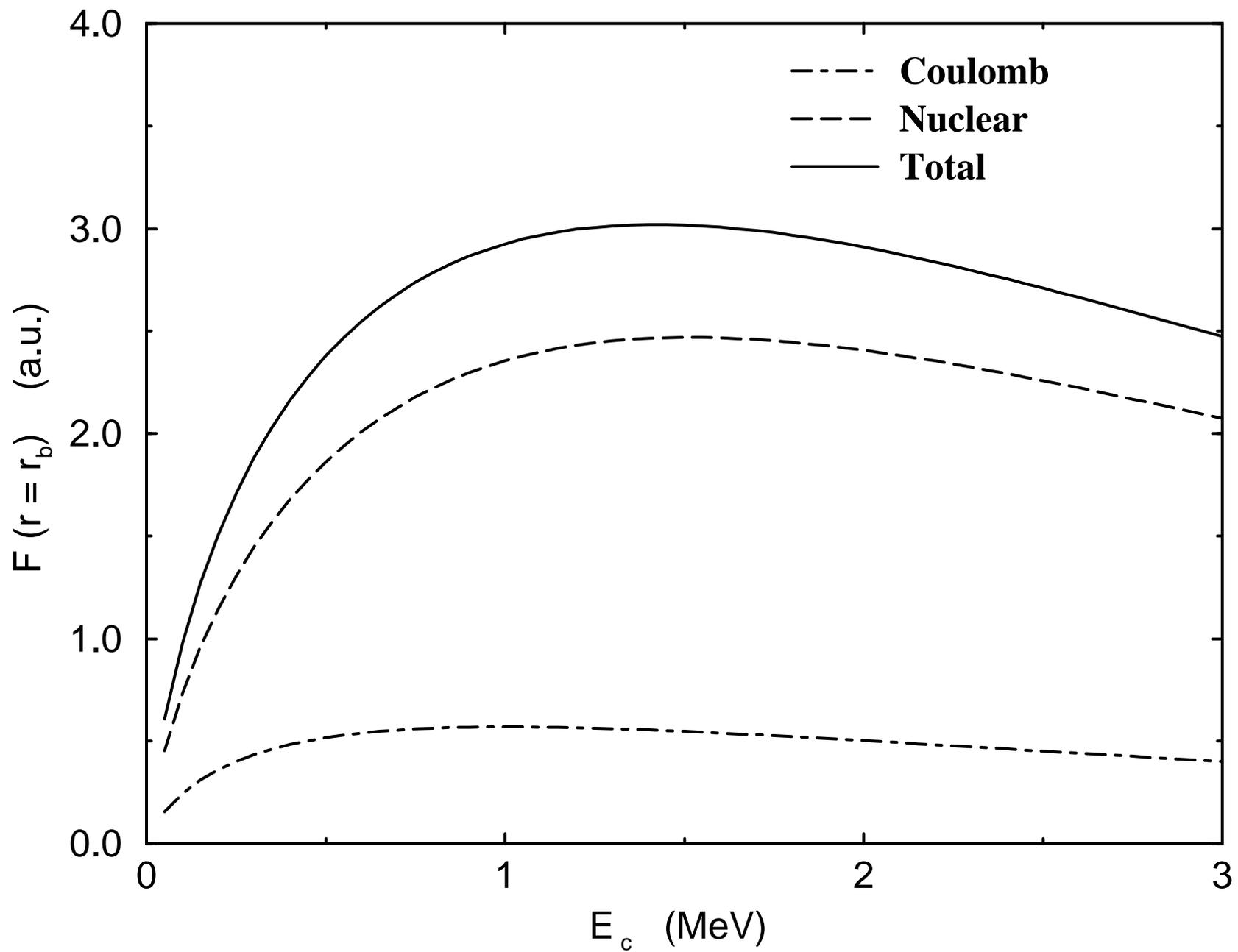
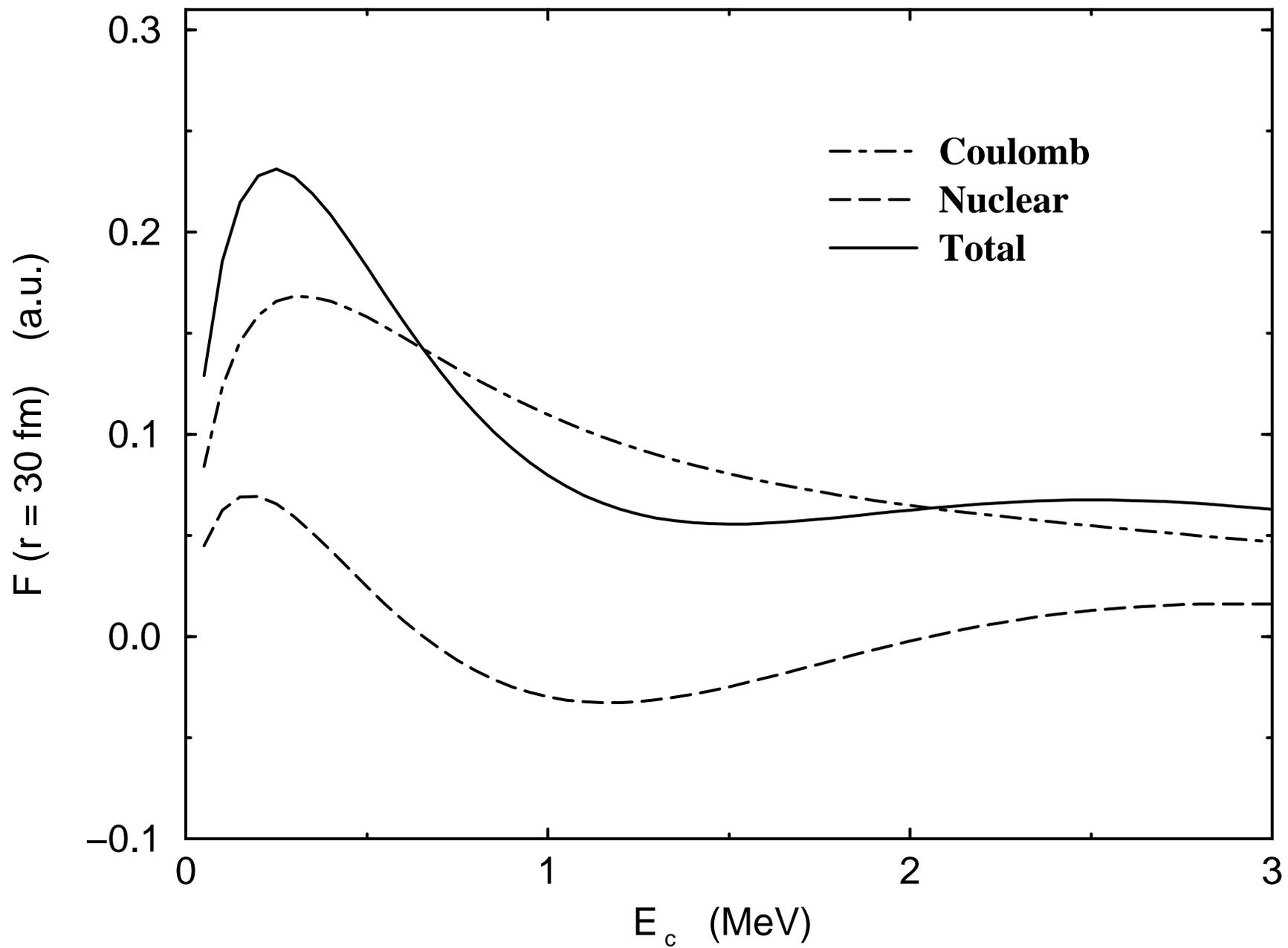
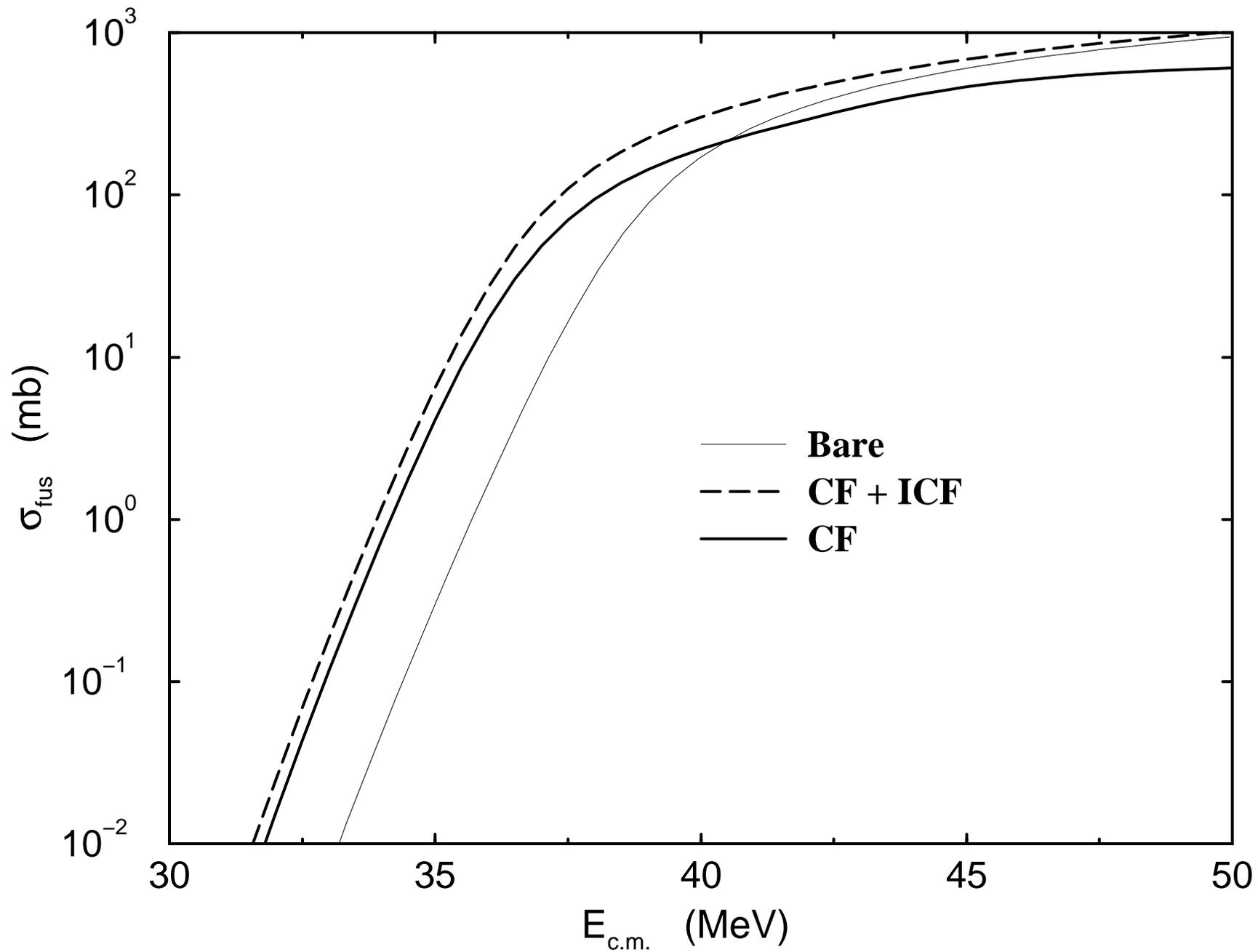
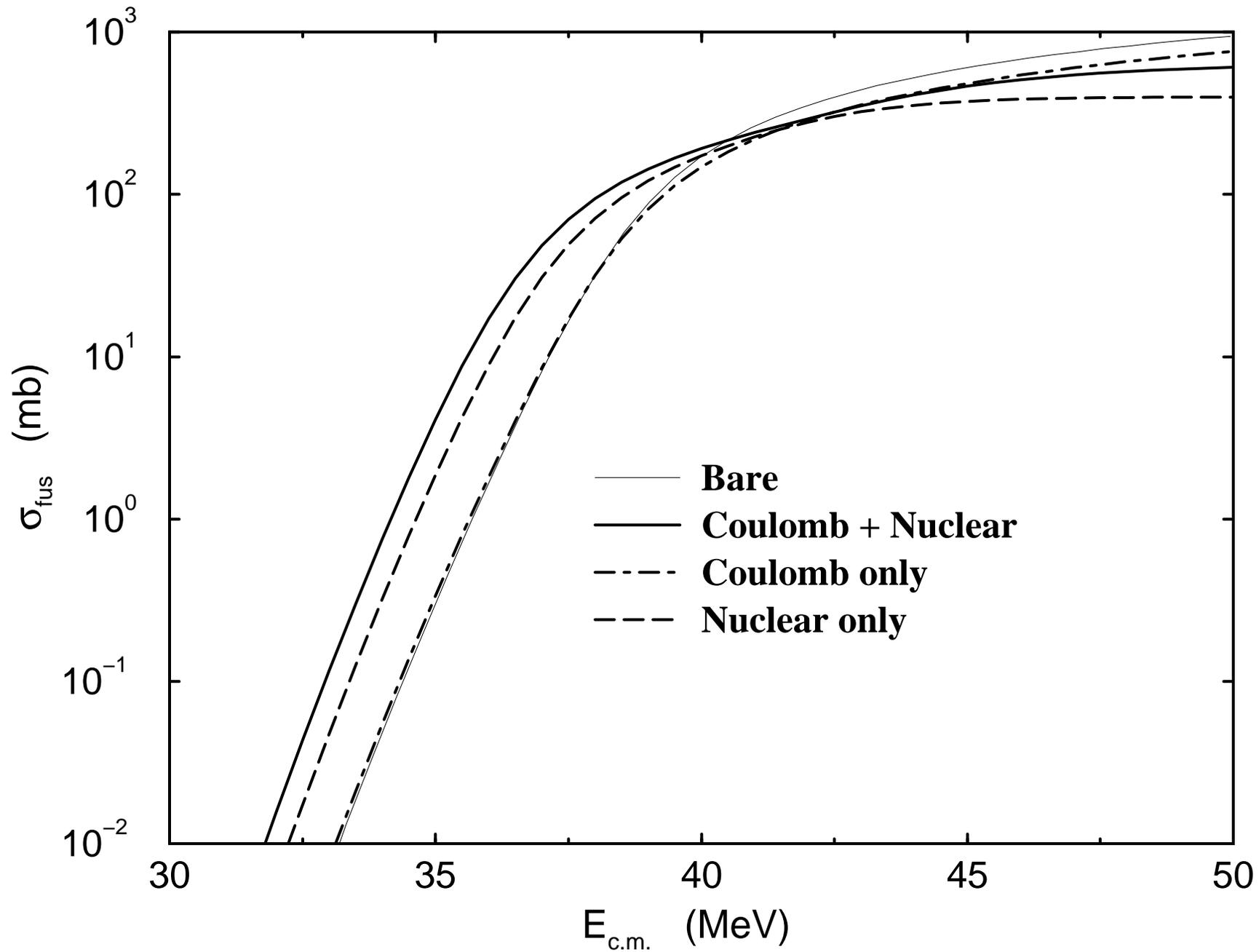
\end{document}